\documentclass[showpacs,twocolumn,aps,superscriptaddress,notitlepage]{revtex4-2}

\usepackage{graphicx}
\usepackage{dcolumn}
\usepackage{bm}
\usepackage{braket}
\usepackage{csquotes}
\usepackage{amsmath,amssymb,amsthm,mathrsfs,amsfonts,dsfont,mathtools}
\usepackage[caption=false]{subfig}
\usepackage{epsfig}
\usepackage{color,xcolor}
\usepackage{adjustbox}
\usepackage{tabularx}
\usepackage{array}
\usepackage{multirow}
\usepackage{tabu}
\usepackage{enumerate}
\usepackage[colorlinks=true,linkcolor=blue,citecolor=blue,urlcolor=blue,pdfauthor={ },pdftitle={ },pdfsubject={ },pdfkeywords={ }]{hyperref}
\usepackage{newtxtext,newtxmath}
\usepackage{makecell} 
\usepackage{url}
\usepackage{booktabs}
\usepackage{nicefrac} 
\usepackage{makeidx}  
\usepackage{leftindex}

\begin{document}


\title{Generative Quantum Data Embeddings for Supervised Learning}

\author{Jaewoong Heo}
 \email{jheo2807@yonsei.ac.kr}
 \affiliation{Department of Statistics and Data Science, Yonsei University, Seoul 03722, Republic of Korea}
\author{Daniel K. Park}
 \email{dkd.park@yonsei.ac.kr}
 \affiliation{Department of Statistics and Data Science, Yonsei University, Seoul 03722, Republic of Korea}
 \affiliation{Department of Applied Statistics, Yonsei University, Seoul 03722, Republic of Korea}
 \affiliation{Department of Quantum Information, Yonsei University, Seoul 03722, Republic of Korea}

\begin{abstract}
Many practically relevant applications of quantum machine learning involve classical data, for which performance depends critically on how inputs are embedded into quantum states. Yet the use of a fixed embedding circuit ansatz remains standard practice.
We propose an energy-based generative learning framework that synthesizes gate sequences to optimize embedding structures and refine data-tailored parameters, using a fidelity-based surrogate objective to guide the search toward improved class distinguishability.
Empirically, the method improves classification performance across diverse settings, while also revealing datasets where architecture search within the present embedding family yields only limited additional gains. We explain this saturation by deriving bounds on the achievable empirical risk in terms of the Wasserstein distance in the input space, showing that classical data geometry provides an \emph{a priori} diagnostic for regimes in which substantial gains from embedding optimization are unlikely. The results establish a practically useful and theoretically motivated framework for searching effective quantum data embeddings through generative optimization, with the attainable gains diagnosed through the geometry of the underlying classical data.
\end{abstract}

\maketitle

\section{Introduction}

Quantum machine learning (QML) has attracted significant interest as a framework for leveraging quantum resources in learning tasks~\cite{biamonte2017quantum,cerezo2022challenges}. While QML is naturally suited to learning tasks involving quantum data~\cite{cerezo2022challenges}, a large class of learning problems of practical interest---ranging from pattern recognition to scientific data analysis---is intrinsically classical~\cite{rebentrost2014quantum,schuld2017implementing,seong2025hamiltonian,choi2025early}. For supervised QML on classical data, model construction typically begins with quantum data embedding, which specifies how classical inputs are encoded into quantum states~\cite{havlivcek2019supervised,giovannetti2008quantum,schuld2019quantum,lloyd2020quantum}. The choice of embedding circuit plays a critical role in determining downstream learning performance, as it fixes the geometry of data in Hilbert space and thereby influences key properties such as expressibility, trainability, and generalization~\cite{schuld2021effect,caro2021encoding,thanasilp2024exponential,hur2025ICML}. In classification tasks, these constraints can be understood through the theory of quantum state discrimination~\cite{helstrom1969quantum,bae2015quantum,giuntini2021quantum}. After embedding, classical data from each class form an ensemble of quantum states, and binary classification can be viewed as the task of choosing a measurement that optimally distinguishes the two class-conditional ensembles in Hilbert space~\cite{bae2015quantum,Lee2024QST}. Since quantum processing cannot increase the distinguishability of these ensembles, the embedding-induced geometry sets a fundamental limit on the class separation available to any downstream classifier~\cite{hur2024neural}. Consequently, supervised learning performance and robustness to noise are constrained by the distinguishability between the embedded class ensembles~\cite{hur2025ICML}.

Despite this central role of data embedding, most QML studies adopt a fixed mapping scheme chosen \emph{a priori}, such as amplitude or angle encoding, typically motivated by practical considerations such as qubit or circuit-depth constraints~\cite{schuld2021machine}. These choices face limitations in near-term QML: amplitude encoding can incur nontrivial state-preparation overhead as the data dimension grows, while simple angle encoding may yield product states and therefore admit efficient classical simulation~\cite{havlivcek2019supervised}. To incorporate more structured inductive bias, Hamiltonian encoding defines the embedding through unitary evolution under a problem-motivated Hamiltonian~\cite{schuld2021machine}. A prominent example is the ZZ feature map (see Appendix~\ref{app:experiment}), an Ising-Hamiltonian inspired construction widely used for its compatibility with noisy intermediate-scale quantum (NISQ) hardware~\cite{Kim2023Nat}, together with the conjecture that repeated applications can induce kernels that are hard to simulate classically~\cite{havlivcek2019supervised}. Nevertheless, such hand-designed feature maps are typically data-agnostic, and their fixed circuit structure can impose an inductive bias that may be misaligned with the geometry of a given dataset, limiting achievable class separation and performance in downstream learning tasks. 

Several methods attempt to inject data dependence without changing the predefined circuit ansatz~\cite{lloyd2020quantum,hubregtsen2022training,gentinetta2023quantum}. One notable example is neural quantum embedding (NQE), which uses a classical neural network to transform the input data, optimized through a quantum fidelity-based loss, and then feeds the transformed data as parameters into a fixed quantum circuit~\cite{hur2024neural,liu2025neural,kim2025multi}. This provides a strong data-dependent baseline that substantially improves separability and classification over standard data-agnostic embeddings. However, despite theoretical and experimental evidence supporting these benefits, such approaches can still be limited by the fixed circuit ansatz. Although the classical preprocessing map is optimized, the quantum embedding remains confined to the family of transformations expressible by the chosen ansatz. Since data-encoding choices determine the accessible function class and induce a specific kernel geometry and inductive bias, the fixed structure may not be well matched to the geometry of the given dataset~\cite{schuld2021effect,caro2021encoding,kubler2021inductive,du2022quantum,incudini2024automatic}.

In this work, we address the limitations of fixed architecture by treating the embedding structure itself as the primary optimization variable. Designing or modifying a quantum circuit architecture to meet a specific objective has been studied in multiple lines of work. Evolutionary strategies search over discrete structures through selection, recombination, and mutation to form a quantum circuit~\cite{pellow2024hybrid,islam2025quantum,phalak2025optimizing}. While this approach is flexible and suited to discrete settings, it may sample candidates inefficiently and thus requires extensive evaluations. Reinforcement learning (RL) has also been used to construct circuits sequentially by choosing circuit elements one by one~\cite{rapp2025reinforcement,fosel2021quantum,kolle2024reinforcement,kundu2024enhancing,montagna2021quantum,altmann2024challenges,ostaszewski2021reinforcement}. Although RL is well-suited to circuit construction, it suffers from a high-variance gradient issue in data-driven settings. Also, the sequential decision process naturally tends to under-represent the synergistic effect that emerges only after a circuit is fully constructed and requires intermediate measurements after each elemental sub-circuit attachment. 

Motivated by these limitations, we adopt a generative modeling strategy in which a quantum circuit is represented as a sequence of words, with each gate or subcircuit treated as a vocabulary token for a large language model~\cite{radford2019language,nakaji2024generative}. In contrast to the previous approaches, our generative embedding architecture search generates complete circuit sequences by sampling from a learned distribution. Since this distribution is updated using feedback from the search objective, later samples are biased toward candidates that are more aligned with the target objective, making the sampling process more goal-directed. Furthermore, because the generative model outputs a full sequence of gates, the sampled circuit can be evaluated after complete construction, avoiding intermediate measurements that arise in step-wise construction. To improve empirical classification performance, we adopt the trace distance between the two embedded class ensembles---a  metric for quantum-state distinguishability---as the guiding objective for the search~\cite{helstrom1969quantum,nielsen2010quantum}. With this data-dependent objective, the generative learning procedure updates its sampling rule to favor candidate circuits with improved objective values. Following this architecture optimization, the quantum data embedding is further refined by introducing a learnable bias function into the gate parameters, thereby exploiting continuous degrees of freedom within the selected circuit structure. The resulting embeddings are evaluated using a quantum kernel support vector machine (QKSVM)~\cite{huang2021power,havlivcek2019supervised} for binary classification, enabling a more direct assessment of the separability induced by the searched embeddings with minimal confounding from the downstream learner. Across multiple datasets, the proposed pipeline yields competitive and, in several cases, superior classification performance relative to fixed-architecture feature maps augmented with NQE.

Beyond constructing effective embeddings, it is also important to understand when embedding optimization can be expected to yield substantial gains. Prior work has characterized performance limits for a fixed embedding map~\cite{helstrom1969quantum}, whereas the present setting produces a family of candidate embeddings whose achievable risk is not captured by fixed-map analysis. To address this, we derive a geometry-based bound on the best empirical loss attainable within the embedding family induced by a given setting (e.g., circuit depth and the available gate set). The bound is expressed in terms of the Wasserstein distance between the class-conditional distributions in the classical input space. When this distance is small, the two classes are weakly separated before quantum encoding, and substantial gains from embedding optimization are unlikely under the given resource setting. The bound therefore serves as a useful screening criterion for identifying datasets where generative embedding optimization is likely to have limited benefit.

Taken together, our contribution is twofold. On the methodological side, we introduce a generative framework for optimizing effective quantum data embeddings beyond fixed circuit templates. On the theoretical side, we show that the attainable gains of such optimization are fundamentally shaped by the geometry of the underlying classical data. This combination yields not only a practical protocol for embedding design, but also an \emph{a priori} diagnostic for when such optimization is unlikely to provide substantial additional benefit.

The remainder of the paper is organized as follows. Section~\ref{sec:II} provides the theoretical background for separability optimization and establishes a dataset-geometry-induced bound on quantum classification performance. Section~\ref{method} describes the generative embedding search procedure and the subsequent continuous parameter refinement step. In Section~\ref{result}, we evaluate the proposed approach using QKSVM, demonstrating classification improvements across multiple datasets and validating the Wasserstein-based interpretation of dataset-dependent saturation.

\section{\label{sec:II}Separability Objective and Geometric Limits}

For a binary classification task, an effective quantum embedding should map classical inputs to quantum states whose class-conditional ensembles are highly distinguishable. This objective has a natural operational meaning through quantum state discrimination and the Helstrom bound. For an embedding structure $\Phi$, the quantum state representation of an input $x$ is $\rho_\Phi(x) = \ket{ \psi_\Phi(x)  }\bra{\psi_\Phi(x)}$. Given a labeled dataset $\{(x_i,y_i)\}_{i=1}^N$ with $y_i\in\{-1,+1\}$, the empirical class priors and class-conditional mixture states can be written as $p^\pm = N_\pm/N$ and $\rho^\pm(\Phi)= \sum_{i:y_i = \pm1} \rho_\Phi(x_i)/N_{\pm}$. Quantum state discrimination then corresponds to the binary decision problem of identifying whether an unknown state is drawn from $\rho^+(\Phi)$ or $\rho^-(\Phi)$ with prior probabilities $p^+$ and $p^-$, while minimizing the error probability over all positive operator-valued measures~\cite{bae2015quantum}. The relevant operational quantity is the trace distance between the prior-weighted class states, $D_{\mathrm{tr}}(p^+\rho^+(\Phi),p^-\rho^-(\Phi))$, which determines the minimum achievable error probability for distinguishing the two hypotheses. Thus, regardless of the particular measurement or classifier used afterward, the empirical risk is bounded as
\begin{equation}
\hat{R}(\Phi) \ge \frac{1}{2} -D_\mathrm{tr}(p^-\rho^-(\Phi), p^+\rho^+(\Phi)), \label{empirical_risk_bound}
\end{equation}
with equality achieved by the Helstrom measurement~\cite{helstrom1969quantum}. This distinguishability cannot be increased by subsequent quantum processing because trace distance is contractive under completely positive trace-preserving maps $\Lambda$, i.e., $D_{\mathrm{tr}}(\Lambda(\rho^+),\Lambda(\rho^-)) \le D_{\mathrm{tr}}(\rho^+,\rho^-)$. Therefore, the trace distance induced by the embedding sets a fundamental limit on the class separation available to any downstream classifier, motivating the optimization of $\Phi$ itself toward larger trace distance~\cite{nielsen2010quantum}.

\subsection{\label{intro:fidelity}Fidelity-based surrogate objective}

Although maximizing the trace distance directly reduces the empirical risk bound in Eq.~(\ref{empirical_risk_bound}), computing it is costly on near-term quantum devices~\cite{wang2023fast}. We therefore use the pairwise fidelity between embedded pure states, $F_\Phi(x_i,x_j)=\left|\braket{\psi_\Phi(x_i)|\psi_\Phi(x_j)}\right|^2$, as a tractable surrogate for trace-distance-based separability. This choice is motivated by the fact that overlaps between pure states can be estimated at substantially lower cost, for example via a swap test~\cite{buhrman2001quantum} or directly from the transition probability~\cite{havlivcek2019supervised}.

The connection between pairwise fidelity and ensemble-level distinguishability can be understood through purity and convexity. The purity of a class-conditional mixture state $\rho^\pm(\Phi)$ is given by
\begin{equation}
\mathrm{Tr}(\rho^\pm(\Phi)^2) = \frac{1}{N^2_\pm} \sum_{i,j:y_i=y_j=\pm1} F_\Phi(x_i, x_j).
\label{purity}
\end{equation}
Thus, increasing within-class fidelities makes each class-conditional mixture more concentrated in Hilbert space. This is desirable because highly mixed class states are intrinsically harder to distinguish. Cross-class fidelities also affect ensemble-level distinguishability through the convexity of the trace norm. In particular,
\begin{equation}
D_\mathrm{tr}(\rho^+(\Phi), \rho^-(\Phi)) \le \frac{1}{N_+N_-} \sum_{y_i\neq y_j} \sqrt{1-F_\Phi(x_i, x_j)}.
\label{convexity}
\end{equation}
Equation~(\ref{convexity}) shows that reducing cross-class fidelities increases the upper bound on the achievable trace distance~\cite{wilde2013quantum,fuchs2002cryptographic}. 

Based on the above considerations, we define a fidelity-based surrogate energy using pairwise labels. For a pair of inputs $x_i$ and $x_j$, the pairwise fidelity loss is
\begin{equation}
\ell_{ij}(\Phi)=|\delta_{y_i y_j}-F_\Phi(x_i, x_j)|
\label{pairwise_fidelity}
\end{equation}
where $\delta_{y_i y_j}=1$ if $y_i=y_j$ and $\delta_{y_i y_j}=0$ otherwise. Minimizing $\ell_{ij}$ promotes large overlaps for same-class pairs and small overlaps for cross-class pairs, providing a practical surrogate objective for embedding search.

Although Eq.~(\ref{pairwise_fidelity}) provides a tractable objective for improving separability, it does not guarantee that the class-conditional mixtures can be made arbitrarily distinguishable. The success of embedding optimization is also constrained by the dataset and by the embedding family under consideration. While previous studies have analyzed classification under a fixed embedding through empirical risk bounds and quantum state discrimination~\cite{hur2024neural,giuntini2021quantum}, it remains less understood how the attainable distinguishability over an embedding family can be bounded \emph{a priori} by the geometry of the classical input data. To address this question, the following subsection develops a Wasserstein-based characterization of the geometric limits on attainable separability.

\subsection{\label{intro:wasser}Wasserstein bounds on empirical loss}
Consider an embedding map structure $\Phi$ composed of parameterized gates drawn from a gate pool $\mathcal{U}$. For each parameterized gate of the form $U(x)=e^{-ixG}$, let $G$ denote its Hermitian generator, and let $\mathcal{G}_\Phi$ be the set of generators associated with the gates appearing in $\Phi$. 
Defining the maximal spectral spread  of $\Phi$ as
$
\Delta_\Phi := \max_{G\in\mathcal{G}_\Phi}(\lambda_{\max}(G)-\lambda_{\min}(G)),
$
where $\lambda_{\max}(G)$ and $\lambda_{\min}(G)$ are the largest and smallest eigenvalues of $G$, respectively, a Lipschitz-type continuity relation between the input data and the embedded quantum states can be written as
\begin{equation}
\| \rho_\Phi(x_i)-\rho_\Phi(x_j)\|_1 \le T_\Phi\Delta_\Phi \|x_i-x_j\|_1.
\label{lipschitz}
\end{equation}
Here $T_\Phi$ denotes the number of data re-uploading layers of $\Phi$~\cite{recio2025single}.

We now extend this single-embedding bound to an embedding family. Consider a circuit-structure search procedure that constructs embedding sequences of length $D$ by selecting depth-one subcircuits from a fixed pool $\mathcal{C}$. Let $\mathcal{F}$ denote the family of embedding structures generated under this fixed sequence length and fixed subcircuit pool. Since any realized circuit has a circuit depth of at most $D$, the number of data re-uploading layers satisfies $T_\Phi\le D$ for all $\Phi \in \mathcal{F}$. In addition, since all parameterized gates in any embedding structure $\Phi \in \mathcal{F}$ are drawn from the fixed pool $\mathcal{C}$, the maximal spectral spread $\Delta_\Phi$ is upper bounded by the family-level maximum $\Delta_\mathcal{F}:=\max_{G\in\mathcal{G}_\mathcal{C}}(\lambda_{\max}(G)-\lambda_{\min}(G))$, where $\mathcal{G}_\mathcal{C}:=\bigcup_{\Phi\in\mathcal{F}}\mathcal{G}_\Phi$ denotes the set of generators represented in the fixed pool $\mathcal{C}$. The continuity bound in Eq.~(\ref{lipschitz}) can then be relaxed uniformly over $\mathcal{F}$ as $\| \rho_\Phi(x_i)-\rho_\Phi(x_j)\|_1 \le D\Delta_\mathcal{F} \|x_i-x_j\|_1$. Defining a new Lipschitz constant $\kappa_\mathcal{F}=D\Delta_\mathcal{F}/2$, a uniform trace distance bound for $\Phi \in \mathcal{F}$ can be derived as

\begin{equation}
D_\mathrm{tr}(\rho_\Phi(x_i), \rho_\Phi(x_j)) \le \kappa_\mathcal{F} \|x_i-x_j\|_1 .\label{trace_lipschitz}
\end{equation}

We next derive a class ensemble-level bound from the per-sample trace-distance bound in Eq.~(\ref{trace_lipschitz}). Let $\Pi(\hat{P}_+(x), \hat{P}_-(x))$ denote the set of all joint distributions (couplings) whose marginals are the empirical class-conditional input distributions $\hat{P}_+(x)$ and $\hat{P}_-(x)$ (see Appendix~\ref{app:proof}). By the convexity of the trace norm and Jensen's inequality~\cite{jensen1906fonctions}, any coupling $\pi \in \Pi(\hat{P}_+(x), \hat{P}_-(x))$ satisfies $D_\mathrm{tr}(\rho^+(\Phi), \rho^-(\Phi)) \le \mathbb{E}_{(x_i, x_j) \sim \pi} \left[ D_\mathrm{tr}(\rho_\Phi(x_i), \rho_\Phi(x_j)) \right]$. Applying the per-sample Lipschitz bound in Eq.~(\ref{trace_lipschitz}) inside the expectation gives
$
D_\mathrm{tr}(\rho^+(\Phi), \rho^-(\Phi))
\le
\kappa_\mathcal{F}
\mathbb{E}_{(x_i,x_j)\sim\pi}
\left[\|x_i-x_j\|_1\right].
$
Since this inequality holds for any coupling $\pi$, we may minimize the right-hand side over all couplings between the two empirical class distributions. By definition, this minimum transport cost is precisely the 1-Wasserstein distance, yielding
\begin{equation}
D_\mathrm{tr}(\rho^+(\Phi), \rho^-(\Phi))
\le
\kappa_\mathcal{F}
W_1(\hat{P}_+(x), \hat{P}_-(x)),
\label{trace_lipschitz_wasser}
\end{equation}
where $W_1$ denotes the 1-Wasserstein distance~\cite{villani2008optimal}.

Assuming balanced class priors, combining Eq.~(\ref{empirical_risk_bound}) with Eq.~(\ref{trace_lipschitz_wasser}) gives a lower bound on the empirical risk attainable by any embedding $\Phi \in \mathcal{F}$
\begin{equation}
\hat{R}^*(X,y;\mathcal{F}) \ge \frac{1}{2} - \frac{1}{2}\kappa_\mathcal{F}W_1(\hat{P}_+(x), \hat{P}_-(x)),
\label{final}
\end{equation}
where $\hat{R}^*(X,y;\mathcal{F})$ denotes the minimum empirical risk achievable over the embedding family when the Helstrom measurement is used. The bound may be loose because of the relaxation $T_\Phi\le D$ and the use of a family-level Lipschitz constant. Nevertheless, it reveals a useful structural dependence: when the empirical class distributions are close in Wasserstein distance, the trace-distance separation attainable by embeddings in $\mathcal{F}$ is limited, making substantial improvements from embedding optimization unlikely under the given resource setting.

\begin{figure}[t]
\centering
\begin{minipage}[t]{0.49\columnwidth}
  \centering
  \includegraphics[width=\linewidth]{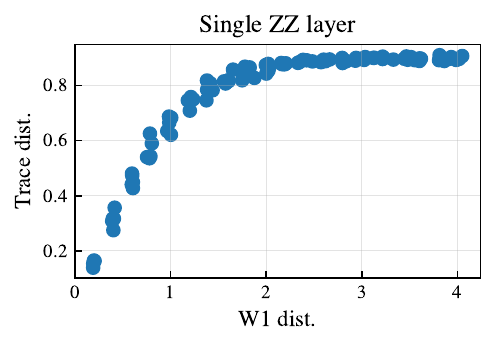}
\end{minipage}\hfill
\begin{minipage}[t]{0.49\columnwidth}
  \centering
  \includegraphics[width=\linewidth]{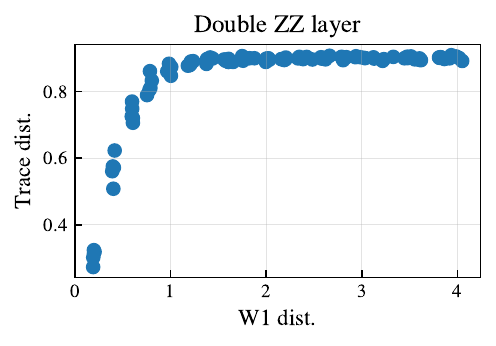}
\end{minipage}
\caption{\label{fig:toy_simul}%
Demonstration of the trace-distance upper bound as a function of the input-space 1-Wasserstein distance. The simulations use a single-layer ZZ feature map (left) and a double-layer ZZ feature map (right). To control the Wasserstein distance between the two classes, the distribution of one class is fixed while the other is spatially translated.}
\end{figure}

Figure~\ref{fig:toy_simul} provides an illustration of the trace distance behavior suggested by Eq.~(\ref{trace_lipschitz_wasser}). As the Wasserstein distance increases, the trace distance between the resulting class-conditional states initially increases and then saturates near its maximal value. Comparing the single-layer and double-layer ZZ feature maps shows that increasing the circuit depth shifts this saturation point toward smaller Wasserstein distances, consistent with the larger family-level Lipschitz constant $\kappa_\mathcal{F}$. These results support the use of input-space Wasserstein distance as a diagnostic for assessing whether substantial gains from embedding architecture search are likely under the given resource setting.

\section{\label{method}Method}
\subsection{\label{method:EGAS}Energy-based generative architecture search}

To search for optimal quantum embedding structures, we propose an energy-based generative architecture search (EGAS) framework. The central idea is to represent a quantum embedding circuit as a discrete sequence and train a generative model to sample circuit structures with low values of a task-specific energy function.

In EGAS, a candidate embedding circuit is represented as a length-$D$ sequence of depth-one subcircuits, $s=(c_1,\dots,c_D)$,
where each token $c_d$ is drawn from a predefined subcircuit pool $\mathcal{C}$. Each token is specified by a tuple $c=(\text{gate type}, \text{qubit index}, \text{data index}, \text{parameter coefficient})$. The gate type is selected from a predefined set of quantum gates, the qubit index specifies the target qubit to which the gate is applied, the data index specifies the classical feature $x_i$ to be injected into the gate, and the parameter coefficient $r$, chosen from a predefined discrete set, specifies the scalar multiplier applied to that feature. Thus, for a parameterized rotation gate, the token defines the data-dependent angle $\phi(x)=r x_i$.
The cardinality of the subcircuit pool, $|\mathcal{C}|$, is determined by the number of valid combinations of these four components.

Each sampled sequence $s$ induces an embedding structure $\Phi_s$, which is evaluated using the pairwise fidelity surrogate introduced in Eq.~(\ref{pairwise_fidelity}). Specifically, we assign to $s$ an energy given by the empirical average of the pairwise fidelity loss over a batch $\mathbb{B}$ of labeled pairs:
\begin{equation}
E(s) = \frac{1}{|\mathbb{B}|} \sum_{b=1}^{|\mathbb{B}|} \ell^b_{ij}(\Phi_s) = \frac{1}{|\mathbb{B}|} \sum_{b=1}^{|\mathbb{B}|} \left| \delta_{y^b_i y^b_j} - F_{\Phi_s}(x^b_i, x^b_j) \right|.
\label{batch_energy}
\end{equation}
By construction, minimizing this energy encourages the embedding to match the desired pairwise fidelity structure: same-class inputs are encouraged to have large quantum-state overlap, whereas cross-class inputs are encouraged to have small overlap. The resulting energy therefore serves as the feedback signal for updating the
generative model. The batch size $|\mathbb{B}|$ controls the trade-off between the statistical stability of this energy estimate and the computational cost of evaluating candidate circuits.

Candidate sequences $s$ are generated using a generative pre-trained transformer (GPT) parameterized by $\theta$. Treating each depth-one subcircuit as a token, the GPT generates the circuit sequence autoregressively: at step $d$, it samples the next token $c_d$ conditioned on the previously generated tokens $(c_1,\dots,c_{d-1})$. The resulting probability distribution over candidate sequences $s=(c_1,\dots,c_D)$ factorizes as $p_\theta(s)=\prod_{d=1}^{D} p_\theta(c_d \mid c_1,\dots,c_{d-1})$. This sequence-modeling formulation provides a flexible distribution over discrete circuit structures and allows the search process to exploit correlations between subcircuits appearing at different depths~\cite{radford2019language}. 

\begin{figure}[h]
\includegraphics[width=\linewidth]{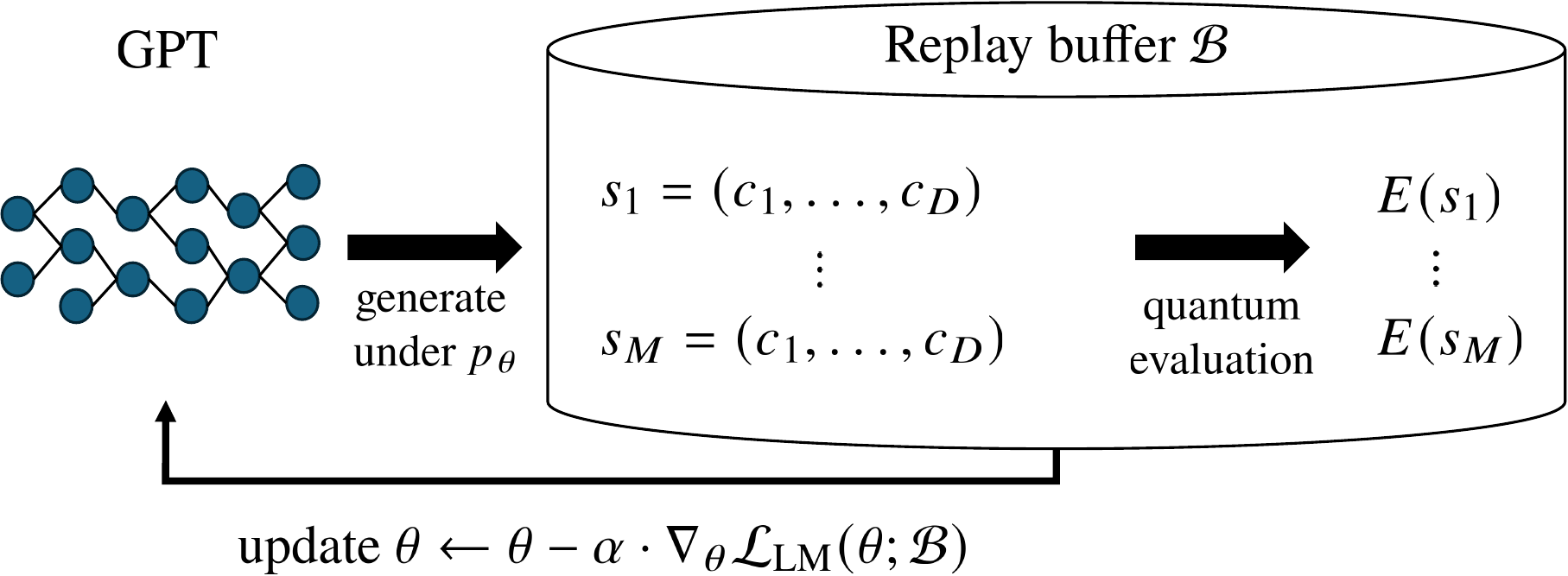}
\caption{\label{fig:egas}Overview of the Energy-based Generative Architecture Search (EGAS) training. In each iteration, a generative pre-trained transformer (GPT) parameterized by $\theta$ autoregressively samples $M$ candidate sequences, $s_1, \dots, s_M$. Each sequence is an ordered length-$D$ list of subcircuit tokens drawn from the predefined pool $\mathcal{C}$. The sequences are translated into quantum embedding circuits and evaluated using a quantum simulator or hardware. This evaluation assigns an energy value $E(s_m)$ to each circuit based on the pairwise fidelity surrogate loss. The evaluated sequence-energy pairs $(s_m,E(s_m))$ are stored in a replay buffer $\mathcal{B}$. During the update phase, the GPT parameters are updated by a gradient step on the logit-matching loss in Eq.~(\ref{logit_matching}) with learning rate $\alpha$.}
\end{figure}

The EGAS training loop is illustrated in Fig.~\ref{fig:egas}. At each iteration, the generator samples $M$ sequences $\{s_m\}_{m=1}^M$ until each sequence reaches the maximum length $D$. Each sequence is then translated into a quantum embedding circuit and evaluated according to Eq.~(\ref{batch_energy}) using either a quantum simulator or quantum hardware. The resulting sequence-energy pairs $(s_m,E(s_m))$ are stored in a replay buffer $\mathcal{B}$ and used to update the generator parameters $\theta$. This replay-based update allows the model to learn from previously evaluated circuits and progressively bias its sampling distribution toward lower-energy regions of the architecture space.

To perform this update, EGAS uses a logit-matching objective that trains the generator to approximate a Boltzmann-like distribution over the evaluated energy landscape. The loss is defined as
\begin{equation}
\mathcal{L}_{\mathrm{LM}}(\theta) = \frac{1}{M} \sum^M_{m=1} \left(e^{-\gamma w_{\mathrm{sum}}(s_m; \theta)}  -   e^{-\gamma E(s_m)}\right)^2 \label{logit_matching}
\end{equation}
where $\gamma > 0$ is an inverse-temperature hyperparameter, and $w_{\mathrm{sum}}(s_m; \theta)$ denotes the cumulative logit score assigned to sequence $s_m$ by the GPT model parameterized by $\theta$. The generator assigns sequence probabilities according to $p_\theta(s) \propto \exp\left(-\gamma w_{\mathrm{sum}}(s; \theta)\right)$. Minimizing Eq.~(\ref{logit_matching}) aligns the model-derived score $w_{\mathrm{sum}}(s_m;\theta)$ with the evaluated energy $E(s_m)$. As a result, sequences with lower evaluated energy are assigned higher sampling probability. Increasing $\gamma$ sharpens this preference and makes the search more strongly concentrated around low-energy circuit structures~\cite{nakaji2024generative}.

\subsection{\label{method:bias}Continuous parameter refinement}
EGAS searches over a finite, tokenized architecture space. Consequently, although it can identify low-energy circuit structures under the surrogate objective, both the circuit architecture and the parameter coefficients are selected from predefined discrete sets. This discreteness is useful for generative architecture search, but it does not exhaust the continuous degrees of freedom available in the corresponding quantum embedding. Since the embedded quantum states vary continuously with gate parameters, a circuit selected by EGAS may still be locally improvable after its discrete structure has been fixed.

To exploit this remaining flexibility, we apply a post-search continuous parameter refinement step. Recall that, for each parameterized gate selected by EGAS, the data-dependent angle has the form $\phi(x)=r x_i$, where the feature index $i$ and coefficient $r$ are fixed by the corresponding token. We then replace this angle with
\begin{equation}
\tilde{\phi}(x)=r x_i + b_\omega(x),
\label{bias_tuning}
\end{equation}
where $b_\omega(x)$ is a parameterized bias function that provides a learnable additive offset. During refinement, the circuit architecture, selected feature index $i$, and base coefficient $r$ are held fixed; only the bias-function parameters $\omega$ are optimized. Thus, the refinement step improves the embedding within a local continuous neighborhood of the EGAS solution without changing the circuit structure discovered during architecture search.

The bias function is initialized as $b_\omega(x)=0$. With this initialization, the refined embedding initially coincides exactly with the embedding selected by EGAS. Any subsequent change in the objective value can therefore be attributed to the continuous adjustment introduced after architecture search, rather than to a different initialization. This separates the roles of the two stages: EGAS selects a discrete circuit structure, whereas continuous parameter refinement determines how much additional improvement can be extracted once that discrete solution is fixed.

For a fixed sequence $s$ selected by EGAS, applying the replacement in Eq.~(\ref{bias_tuning}) to the embedding $\Phi_s$ yields a bias-refined embedding $\Phi_{s,\omega}$. The bias function parameters $\omega$ are optimized using a binary cross-entropy loss over a batch $\mathbb{B}$ of paired inputs, defined as
\begin{equation}
\begin{split}
\mathcal{L}_{\mathrm{bias}}(\omega;s) &= -\frac{1}{|\mathbb{B}|} \sum_{b=1}^{|\mathbb{B}|} \bigg[ \delta_{y_i^b y_j^b} \log \bar{F}_{\Phi_{s,\omega}}(x_i^b,x_j^b) \\
&\quad + \left(1-\delta_{y_i^b y_j^b}\right) \log\left(1-\bar{F}_{\Phi_{s,\omega}}(x_i^b,x_j^b)\right) \bigg],
\end{split}
\label{BCE_loss}
\end{equation}
where $\bar{F}_{\Phi_{s,\omega}}(x_i^b,x_j^b)
=
\min\{\max\{F_{\Phi_{s,\omega}}(x_i^b,x_j^b),\epsilon\}, 1-\epsilon\}$ represents the clipped pairwise fidelity under the embedding $\Phi_{s,\omega}$, clamped with a small threshold $\epsilon>0$ to ensure numerical stability by preventing the logarithmic terms from becoming undefined at $0$ or $1$. 

\section{\label{result}Results}

\subsection{\label{result:bias_anal}Empirical impact of bias refinement}

We first examine the empirical effect of continuous parameter refinement after EGAS. To this end, the EGAS outputs are partitioned according to their pre-refinement surrogate energy $E$. After removing duplicate sequences, we retain the ten unique sequences with the lowest energies, denoted by $\{G_k\}_{k=1}^{10}$, and the ten unique sequences with the highest energies, denoted by $\{B_k\}_{k=1}^{10}$.

\begin{figure}[t]
\includegraphics[width=\linewidth]{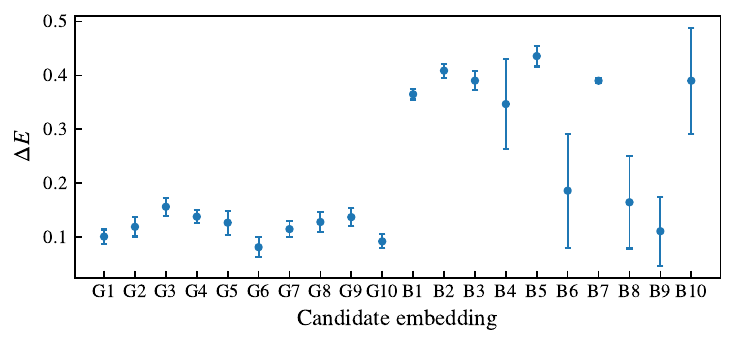}
\caption{\label{fig:delta_by_circ} Surrogate energy reduction ($\Delta E$) achieved by continuous bias refinement on the Phishing Website dataset. The x-axis labels $G$ and $B$ denote the ten unique lowest-energy and ten unique highest-energy circuit architectures identified during EGAS, respectively. Each candidate embedding was refined ten times. Blue circles represent the mean energy reduction, and error bars indicate the standard deviation over repeated runs.}
\end{figure}

Because refinement is performed with the circuit architecture fixed, the quantity $\Delta E = E_{\mathrm{before}} - E_{\mathrm{after}}$ isolates the reduction in surrogate energy attributable to continuous parameter refinement. Figure~\ref{fig:delta_by_circ} reports this quantity for the Phishing Website (PW)~\cite{phishing_websites_327} dataset over the 20 selected candidate circuits. Experimental details are provided in Appendix~\ref{app:experiment}.

As shown in Fig.~\ref{fig:delta_by_circ}, continuous parameter refinement reduces the surrogate energy for all selected candidate structures, irrespective of their initial assignment to the $G$ or $B$ group. The reductions are generally modest and stable for the initially low-energy $G$ candidates, whereas several initially high-energy $B$ candidates exhibit larger reductions, often accompanied by larger variability across repeated runs. This indicates that architectures obtained through discrete search can retain locally exploitable degrees of freedom in the continuous parameter space, and that the extent of this remaining flexibility depends strongly on the selected circuit. Overall, these results support the role of continuous parameter refinement as a complementary step to EGAS: the discrete search identifies candidate circuit structures, while refinement further adjusts the embedding within the local continuous neighborhood of each selected structure.

\begin{figure}[t]
\includegraphics[width=\linewidth]{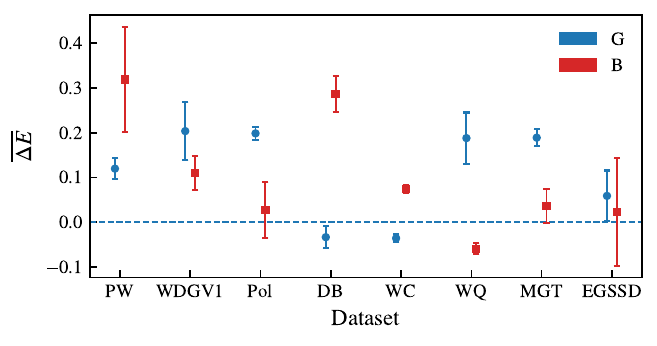}
\caption{\label{fig:delta_by_data} Continuous parameter refinement across eight datasets: PW, WDGV1, Pol, DB, WC, WQ, MGT, and EGSSD. The y-axis shows the mean surrogate energy reduction $\overline{\Delta E}$. For each dataset and group, ten embeddings were considered, each with ten repetitions, and the summary statistics are plotted. Blue circles and error bars represent the mean and standard deviation for group $G$, and red squares and error bars represent the mean and standard deviation for group $B$. The dashed line at $y=0$ is used to indicate the sign of the bias refinement effect.}
\end{figure}

Figure~\ref{fig:delta_by_data} extends this analysis by reporting the group-wise mean energy reduction, $\overline{\Delta E}$, for the $G$ and $B$ groups within each dataset. The analysis is performed across eight datasets: Phishing Website (PW)~\cite{phishing_websites_327}, Waveform Database Generator Version 1 (WDGV1)~\cite{waveform_database_generator_version_1_107}, Pol~\cite{vanschoren2014openml, grinsztajn2022tree}, Dry Bean (DB)~\cite{dry_bean_602}, Wine Color (WC)~\cite{wine_quality_186}, Wine Quality (WQ)~\cite{wine_quality_186}, Magic Gamma Telescope (MGT)~\cite{magic_gamma_telescope_159}, and Electrical Grid Stability Simulated Data (EGSSD)~\cite{electrical_grid_stability_simulated_data__471}. The cross-dataset results show that continuous refinement frequently reduces the surrogate energy, while the magnitude and group-wise pattern of the reduction remain strongly dataset-dependent. In particular, the comparison between the initially low-energy $G$ group and the initially high-energy $B$ group does not exhibit a universal ordering. Some datasets exhibit larger average reductions for the $B$ group, whereas others show comparable or even larger reductions for the $G$ group.

Figure~\ref{fig:delta_by_data} also shows that bias refinement does not guarantee a positive average $\Delta E$ for every dataset and group. In datasets such as DB and WC, the mean reduction is close to zero or slightly negative. This does not necessarily indicate a failure of refinement; rather, it suggests that the selected architectures may already lie near low-energy regions, leaving limited room for further local improvement. In such cases, any additional gain from bias adjustment can become comparable to run-to-run variability in the refinement process, leading to a negligible average change.

\subsection{\label{result:svm_result}Embedding evaluation via quantum kernel SVM}

The preceding analysis shows that continuous parameter refinement can reduce the surrogate energy in many cases. However, the practical value of this reduction must be assessed through downstream classification performance. We therefore evaluate the resulting embeddings using a QKSVM, comparing architecture-only and architecture-plus-refinement embeddings.

To assess whether the optimized separability translates into downstream performance, the embeddings are evaluated through binary classification using a QKSVM. This choice is natural because each embedding $\Phi$ directly defines a quantum kernel through the fidelity between embedded states, $K_{ij}=F_\Phi(x_i,x_j)$. A support vector machine (SVM) provides a standard supervised learner for such precomputed kernels. It uses the kernel matrix to construct a decision boundary in the embedding-induced feature space, without requiring an additional trainable quantum model or a highly flexible downstream neural network. As a result, the QKSVM evaluation focuses on the quality of the embedding-induced kernel geometry itself.

To account for the dependence of downstream accuracy on the particular train-test partition, each embedding is evaluated over ten train-test splits. Within each split, all embeddings are evaluated on the same training and test samples, so that their accuracies can be compared under a common evaluation condition. The classical baseline is a linear SVM trained on the standardized classical input features. This provides a direct reference for linear class separability in the input feature space. A nonlinear RBF-SVM result is also reported in the Appendix~\ref{app:heatmap}. For each split, the QKSVM accuracy of an embedding is compared with the accuracy of this classical baseline, producing a win, tie, or loss.

\begin{figure}[t]
\includegraphics[width=\linewidth]{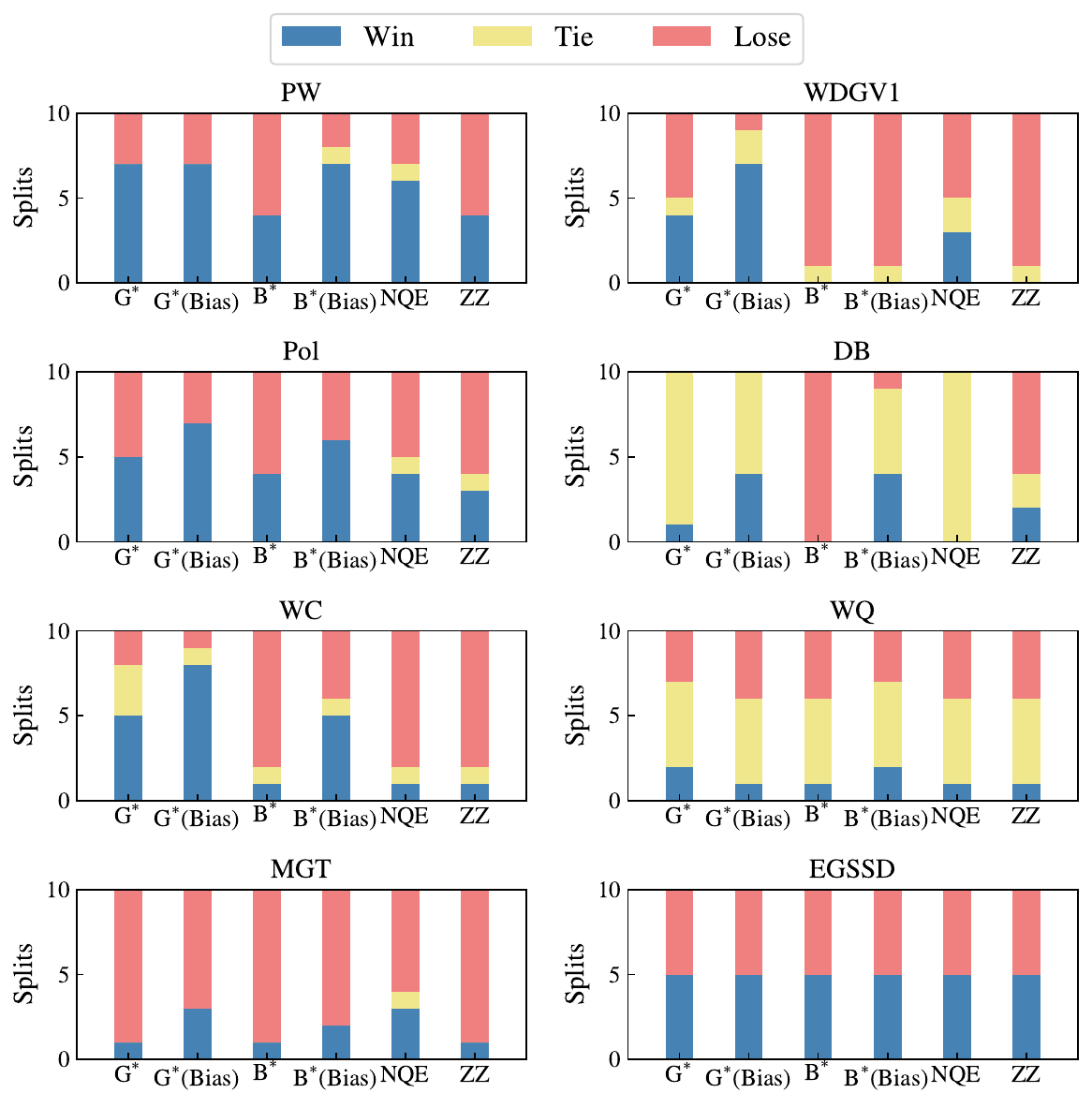}
\caption{\label{fig:winning_summary} Split-wise comparison against the classical baseline across datasets. For each dataset, stacked bars report the numbers of wins (blue), ties (yellow), and losses (red) over ten test splits relative to the classical linear SVM. For each EGAS-derived group [$G$, $G(\mathrm{Bias})$, $B$, and $B(\mathrm{Bias})$], the symbol $*$ denotes the selected candidate embedding that attains the largest number of wins over the classical baseline for that dataset. Here, ZZ denotes the vanilla ZZ feature map, while NQE denotes the same ZZ feature map augmented with trainable neural preprocessing. Because both baselines use a fixed circuit structure, they appear as single reference models rather than as families of sampled architectures.}
\end{figure}

Figure~\ref{fig:winning_summary} visualizes this split-wise comparison. Each EGAS-generated group, $G$, $G(\mathrm{Bias})$, $B$, and $B(\mathrm{Bias})$, contains multiple candidate embeddings. To summarize each group by its strongest representative, we report the embedding that achieves the largest number of wins against the classical baseline across the ten train-test splits. The symbol $*$ marks this selected representative. The ZZ and NQE baselines use fixed circuit structures and therefore appear as single reference models.

This split-wise summary shows how consistently each selected model outperforms the classical linear SVM across the ten train-test splits. A large number of wins suggests that the observed advantage is not limited to a particularly favorable split, whereas ties or losses indicate that the advantage is weak or unstable across different train-test partitions. Across the datasets, the selected representative from the $G(\mathrm{Bias})$ group most frequently achieves the strongest performance. The data-agnostic ZZ feature map is often among the weakest models, while the NQE baseline generally occupies an intermediate position: it improves over ZZ in many cases but does not consistently match the best EGAS-derived architectures.

The dataset-specific results further clarify the role of continuous parameter refinement. In several datasets, especially PW, Pol, DB, and WC, the refined low-energy group $G(\mathrm{Bias})$ achieves one of the largest win counts. This suggests that continuous refinement can effectively improve embeddings when it is applied to circuit architectures that already have favorable surrogate energy. The PW dataset also shows that refinement can sometimes improve initially high-energy architectures: although the selected $B$ embedding yields few wins before refinement, the selected $B(\mathrm{Bias})$ embedding becomes comparable to the selected $G(\mathrm{Bias})$ embedding after refinement.

At the same time, the results show that refinement cannot compensate for every unfavorable architecture. In WDGV1, the selected $B(\mathrm{Bias})$ embedding remains weak even after refinement, indicating that local continuous adjustment is not sufficient when the underlying circuit structure is poorly suited to the task. Thus, continuous parameter refinement should be viewed as a complementary step to architecture search, rather than as a substitute for selecting an appropriate circuit structure.

\begin{figure}[t]
\includegraphics[width=\linewidth]{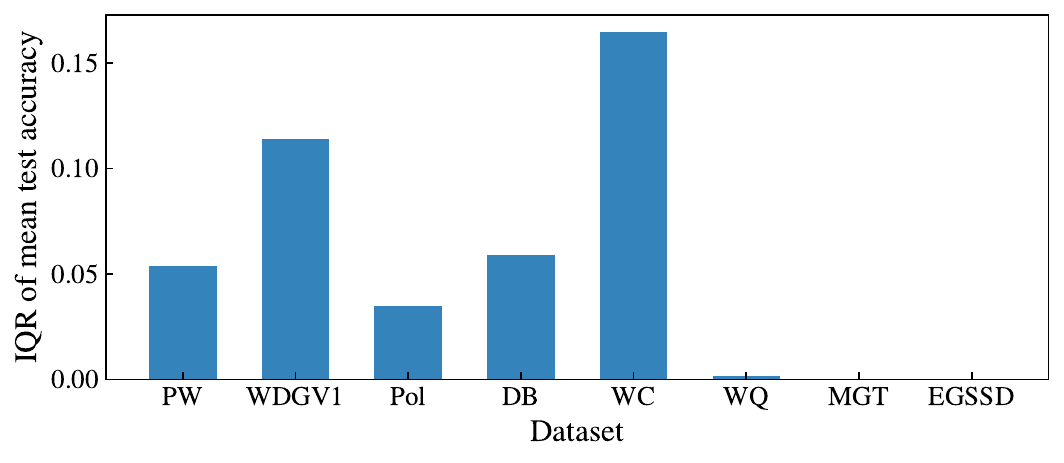}
\caption{\label{fig:dispersion} Dataset-level dispersion of QKSVM test accuracy across the evaluated quantum embeddings. For each evaluated embedding, we first compute the mean test accuracy over the ten train-test splits. For each dataset, the bar height gives the interquartile range (IQR) of these mean accuracies over the fixed quantum baselines (ZZ and NQE) and the EGAS-derived candidate embeddings from the $G$ and $B$ groups, both before and after bias refinement. The classical linear SVM baseline is excluded from this dispersion. A larger IQR indicates stronger variation in classification accuracy across the evaluated embeddings, whereas a smaller IQR indicates relative insensitivity to the embedding choice.}
\end{figure}

A different pattern appears for WQ, MGT, and EGSSD. In these datasets, the overall winning summary is comparatively weak, and $G(\mathrm{Bias})$ does not exhibit a dominant advantage. For WQ, none of the selected EGAS-derived embeddings shows a clear advantage over the classical baseline, and the split-wise outcomes are dominated by ties. For MGT, the selected EGAS-derived embeddings lose to the classical baseline on several splits, including the representative from $G(\mathrm{Bias})$, indicating that refinement does not yield a clearly competitive embedding for this dataset. For EGSSD, all evaluated models exhibit nearly identical win-tie-loss patterns, suggesting that QKSVM performance is only weakly sensitive to the choice of embedding. These observations motivate the embedding-sensitivity analysis in Fig.~\ref{fig:dispersion}, which quantifies how strongly downstream accuracy varies across the evaluated embeddings for each dataset.

Figure~\ref{fig:dispersion} quantifies, for each dataset, how sensitive downstream classification performance is to the choice of quantum embedding within the evaluated set. To obtain this dispersion, we first compute the mean test accuracy for each individual quantum embedding, averaged over the ten repeated train-test splits. These mean accuracies form a dataset-specific distribution consisting of the fixed quantum baselines (ZZ and NQE), and all EGAS-derived candidate embeddings from the $G$ and $B$ groups, both before and after bias refinement. The classical linear SVM baseline is excluded from this distribution because it does not correspond to a quantum embedding and is used only as an external reference in the split-wise comparison. The interquartile range (IQR), defined as the difference between the $75$th and $25$th percentiles of this distribution, is then used as a compact measure of embedding sensitivity. A large IQR indicates that different embeddings yield substantially different mean test accuracies, whereas a small IQR indicates that the evaluated embeddings are concentrated within a narrow performance range.

This IQR analysis makes the dataset-wise contrast more explicit. PW, WDGV1, Pol, DB, and WC show appreciable dispersion, indicating that downstream accuracy remains sensitive to the embedding choice. By contrast, WQ, MGT, and EGSSD have markedly smaller IQR values. For these datasets, the evaluated embeddings are concentrated within a narrow accuracy range, suggesting that the weak win-tie-loss patterns in Fig.~\ref{fig:winning_summary} are not merely due to the absence of a favorable representative. Rather, these datasets exhibit limited performance differentiation across the evaluated embeddings.

Appendix~\ref{app:heatmap} further supports this interpretation by reporting the kernel-level mean test accuracy over the ten train-test splits for all evaluated embeddings. In WQ, MGT, and EGSSD, the accuracies remain tightly clustered across all evaluated embeddings, consistent with their small IQR values in Fig.~\ref{fig:dispersion}. This behavior contrasts with datasets such as PW, WDGV1, DB, and WC, where the heatmap shows a wider spread of QKSVM accuracies across the evaluated models. The weak split-wise behavior in Fig.~\ref{fig:winning_summary} for WQ, MGT, and EGSSD should therefore not be interpreted as an artifact of selecting a particular embedding, but rather as evidence that the QKSVM test accuracy is relatively insensitive to embedding choice within the evaluated embedding set.

\begin{table}[t]
\caption{\label{tab:wasserstein_dist}%
Empirical 1-Wasserstein distances between the class-conditional input distributions for each dataset.
}
\begin{ruledtabular}
\begin{tabular}{lr}
\textrm{Dataset} & \multicolumn{1}{c}{\textrm{1-Wasserstein dist.}}\\
\colrule
PW & 5.2380\\
WDGV1 & 5.1570 \\
Pol & 4.3543 \\
DB & 13.9108 \\
WC & 10.8562\\
\textbf{WQ} & 3.0112\\
\textbf{MGT} & 3.3036\\
\textbf{EGSSD} & 3.5619\\
\end{tabular}
\end{ruledtabular}
\end{table}

Table~\ref{tab:wasserstein_dist} reports the empirical 1-Wasserstein distance between the class-conditional input distributions for each dataset. These distances provide an input-space geometric diagnostic for interpreting the observed QKSVM performance trends. The three datasets with the smallest Wasserstein distances are WQ, MGT, and EGSSD, which are also the datasets with weak variation in QKSVM accuracy across the evaluated embeddings and small IQR values in Fig.~\ref{fig:dispersion}. This agreement is consistent with the Wasserstein-based discussion in Section~\ref{intro:wasser}: when the class-conditional input distributions are weakly separated, the trace-distance separation attainable within a fixed embedding family can be limited, making large architecture-dependent gains less likely.

By contrast, datasets such as WC and DB exhibit broader variation in QKSVM accuracy across the evaluated embeddings. Their IQR values are substantially larger, and the heatmap in Appendix~\ref{app:heatmap} shows a wider spread of QKSVM accuracies across the evaluated embeddings. For these datasets, the choice of embedding can therefore have a clear effect on classification performance, and the EGAS-derived embeddings yield candidate embedding that outperform the classical baseline on multiple train-test splits. Since the same QKSVM protocol is used throughout, this difference in sensitivity is not attributable to classifier flexibility. Instead, it is consistent with the Wasserstein-based interpretation that datasets with larger input-space class separation can allow larger performance differences among embeddings within the explored embedding family.

\section{Conclusion}

In this work, we introduced EGAS, an energy-based generative architecture search framework for optimizing quantum data embeddings in supervised learning. The proposed approach formulates the design of quantum data embeddings as an energy-based optimization problem, where candidate circuit structures are evaluated using a pairwise-fidelity surrogate for class separability. A generative model is then trained to explore the combinatorial space of discrete circuit architectures and to bias the search toward lower-energy embeddings. To complement this discrete architecture search, we further introduced a continuous parameter refinement step that locally adjusts the embedding through a learnable bias function while preserving the circuit structure selected by the generative model.

Our empirical results show that EGAS consistently identifies embeddings whose kernel-SVM performance matches or exceeds that of the existing quantum embedding baselines, including ZZ and NQE, across all evaluated datasets. The comparison between before and after continuous bias refinement demonstrates that continuous refinement can provide additional gains after discrete search, but also that its effect is strongly architecture- and dataset-dependent. In most evaluated datasets, the selected EGAS-derived embeddings also outperform the classical baseline across multiple train-test splits. In other datasets, however, the evaluated embeddings lead to tightly clustered QKSVM accuracies, indicating that the available search space provides only limited variation in classification accuracy under the fixed depth and gate-pool setting considered here.

To interpret this dataset dependence, we developed a Wasserstein-based geometric perspective on the limits of embedding optimization. The resulting bound connects the distinguishability attainable by an embedding family to the 1-Wasserstein distance between class-conditional input distributions. This analysis provides an \emph{a priori} diagnostic for identifying datasets in which large architecture-dependent gains are unlikely within a fixed embedding family. Empirically, datasets with small Wasserstein distances exhibit weak variation in QKSVM performance across the evaluated embeddings, whereas datasets with larger input-space class separation show stronger sensitivity to embedding choice. These observations suggest that the success of quantum embedding search is governed not only by the expressivity of the circuit family, but also by the geometry of the underlying classical data.

Taken together, these results establish a framework in which quantum data embeddings can be optimized through generative architecture search, refined through local continuous parameter adjustment, and interpreted through input-space geometry. The method provides a practical route for discovering task-adapted quantum embeddings, while the Wasserstein-based analysis offers a complementary diagnostic for assessing when such optimization is likely to be beneficial. 

The present framework also opens several directions for future work. First, the current pipeline separates discrete architecture search from continuous parameter refinement. As a result, architectures that appear suboptimal under discretized coefficients may be assigned low priority, even though they could become effective after joint continuous tuning. A natural extension is to couple structural and parametric optimization more tightly, for example through alternating updates, differentiable relaxations of architectural choices, or generative models that propose both circuit structure and continuous parameters~\cite{furrutter2025synthesis}. Second, the energy used to guide architecture search is estimated from a finite batch of sampled training pairs. Relying on such a fixed training batch may bias the ranking of candidate embeddings toward the specific pairs used during the search. This issue could be addressed by resampling-based selection procedures~\cite{efron1992bootstrap} or curriculum-style schedules that gradually increase the diversity and difficulty of the sampled pairs~\cite{bengio2009curriculum}. Finally, deployment on near-term quantum hardware will require search spaces tailored to hardware-native gate sets and connectivity constraints, together with a systematic analysis of how the trade-off between circuit depth and noise affects classification performance.

\begin{acknowledgments}
This work was supported by the Yonsei University Research Grant of 2025, Institute of Information \& communications Technology Planning \& evaluation (IITP) grant funded by the Korea government (No. 2019-0-00003, Research and Development of Core Technologies for Programming, Running, Implementing and Validating of Fault-Tolerant Quantum Computing System), the National Research Foundation of Korea (RS-2025-02309510), the Ministry of Trade, Industry, and Energy (MOTIE), Korea, under the Industrial Innovation Infrastructure Development Project (RS-2024-00466693), and Korean ARPA-H Project through the Korea Health Industry Development Institute (KHIDI), funded by the Ministry of Health \& Welfare, Korea (RS-2025-25456722).
\end{acknowledgments}

\appendix

\section{Experimental details}
\label{app:experiment}
Experiments are conducted on multiple datasets, each converted into a binary classification task by selecting two classes with sufficient numbers of data points. The dimension of input features are reduced to match the number of qubits $n$ using a fixed principal component analysis pipeline, and all reported results are obtained with $n=8$. The range of reduced feature vector values is rescaled to the range $[0, 2\pi]$ prior to embedding so that the features can be injected naturally into rotation-based gates. The training data used for architecture search and bias refinement are converted into pairwise inputs $(x_i, x_j)$ with Kronecker delta target $\delta_{y_i, y_j} \in \{0,1\}$, where the true labels are $y \in \{-1, +1\}$. This paired construction allows the fidelity surrogate objective to train the embedding to predict whether two samples come from the same class through the overlap $F_\Phi(x_i,x_j)$. All reported metrics are averaged over repeated runs with independent random seeds. Quantum simulation is carried out with PennyLane's \texttt{lightning.qubit} backend using analytic probabilities (\texttt{shots=None}) so that the reported differences between embeddings are not affected by finite-shot measurement noise~\cite{bergholm2018pennylane}.

All embedding circuits are represented as sequences of depth-one sub-circuit tokens drawn from a predefined pool $\mathcal{C}$. The pool includes single-qubit parameterized gates $\{\text{RX}, \text{RY}, \text{RZ}  \}$, single-qubit non-parameterized gates $\{\text{H},\text{I}\}$, and two-qubit entangling gates $\{\text{CNOT}, \text{MultiRZ}\}$. This pool is intentionally constructed to balance expressivity with search efficiency. Specifically, the parameterized set $\{\text{RX}, \text{RY}, \text{RZ}  \}$ ensures full single-qubit rotational freedom, while $\text{CNOT}$ serves as the standard two-qubit entangling basis. The identity gate $\text{I}$ acts as a no-op, providing the generative model with the flexibility to effectively search shallower circuits by omitting operations. Furthermore, $\{\text{H}, \text{MultiRZ}\}$ are included as structural shortcuts to accelerate the search process by providing immediate access to these widely used quantum operations.

Each parameterized sub-circuit token additionally specifies a data-feature index and a coefficient chosen from the coefficient grid $r \in \{0.1, 0.3, 0.5, 0.7, 1.0  \}$. A token therefore corresponds to a combination of gate type, qubit index, feature index, and coefficient. The maximum sequence length is fixed to $D=28$, matching the depth of the single-layer ZZ feature map baseline. In this implementation, the baseline is written as
\begin{equation}
U_\mathrm{ZZ}(x) = \exp \left(  i \sum^n_{j=1} \vartheta_j(x)Z_j +   i \sum^n_{j=1} \vartheta_{j, j+1}(x)Z_j Z_{j+1}    \right) H^{\otimes n},
\label{zzfeaturemap}
\end{equation}
where $\vartheta_j(x) = x_j$ and $\vartheta_{j,j+1}(x) = (\pi-x_j)(\pi-x_{j+1})$.

\subsection{Details on architecture search}

Architecture search is performed using an autoregressive GPT generator with vocabulary size $|\mathcal{C}|+1$, including a start token. Training proceeds for $4,000$ iterations, with a linearly decreasing temperature schedule from $\mathcal{T}_{\max}=100$ to $\mathcal{T}_{\min}=0.04$. At each iteration, candidate sequences are generated and evaluated using the fidelity-surrogate energy. To stabilize training, the energies are normalized using exponential moving estimates of the mean and standard deviation, after which a subset of sequences is selected for model updates. The selected subset consists of the lowest-energy and highest-energy fractions $k$, together with an additional middle-energy fraction $k/2$ drawn from the remaining sequences. The selected sequences are randomly permuted before batching. This top-middle-bottom selection strategy ensures that while low-energy sequences guide the search toward promising candidates, high-energy sequences provide explicit contrastive pressure. This prevents the logit-to-energy regression from collapsing into a narrow energy band and helps the generator distinguish candidate sequences across a broader range of energies. Including the middle ranks further stabilizes training by maintaining diversity and mitigating overfitting to extreme candidates. The optimization uses Adam with a learning rate of $5 \times 10^{-5}$, a weight decay of $10^{-2}$, and momentum parameters $(\beta_1, \beta_2)=(0.9, 0.999)$.

\subsection{Details on bias tuning}

The bias function $b_\omega(x)$, which provides an additive offset to all parameterized gates, is implemented as a small multilayer perceptron with a zero-initialized output head. Its output is scaled by a constant gain factor of $10$ before being injected into the circuit parameters. Training uses RMSprop with learning rate $5\times 10^{-4}$ and gradient clipping with maximum norm $2.0$. Each refinement run is trained for $400$ epochs with minibatches of size $25$. A small $L_2$ penalty with coefficient $10^{-6}$ is applied to the output of the bias function $b_\omega(x)$ to discourage unnecessarily large additive shifts. The final reported refinement metrics are averaged over the last $10$ epochs to reduce sensitivity to late-epoch fluctuations.

\subsection{Details on quantum kernel SVM}
Downstream evaluation is performed with a QKSVM to isolate the effect of embedding geometry. For each train-test split, an SVM is trained with a precomputed quantum kernel: the training kernel matrix $K_\mathrm{train}$ is constructed from all train-train pairs, and the test kernel matrix $K_\mathrm{test}$ is constructed from all test-train pairs. A fixed regularization parameter $C=0.05$ is applied across all evaluated embeddings. In addition to the QKSVM evaluation of quantum embeddings, a classical baseline is evaluated using a linear SVM trained on standardized input features. The $z$-scoring statistics for this standardization are computed strictly from the training split, and the same regularization parameter $C=0.05$ is used. For the heatmap in Appendix~\ref{app:heatmap}, we additionally report an RBF-kernel SVM trained on the same standardized input features. The regularization parameter is kept fixed at $C=0.05$, and the RBF kernel coefficient is fixed to $0.125$ for all datasets.

Repeated evaluation is performed over a sequence of non-overlapping data slices consisting of $400$ training samples and $50$ test samples, with the slice start index shifted across repeats. The procedure is repeated ten times, and within each repeat all embeddings are evaluated on exactly the same split. Under this protocol, performance variability reflects not only optimization randomness but also changes in the sampled test population.

\section{Wasserstein-based empirical risk bounds for embedding families}
\label{app:proof}

Let

\begin{gather*}
S := \{(x_i,y_i)\}_{i=1}^N,\qquad y_i\in\{-1,+1\} \\
p^\pm := \frac{N^\pm}{N}, \qquad N^\pm := \#\{i:\,y_i=\pm1\}. \\
\hat P_\pm(x) := \frac{1}{N^\pm}\sum_{i:\,y_i=\pm1}\delta_{x_i}(x). \\
\rho^\pm(\Phi) := \mathbb{E}_{x\sim \hat P_\pm(x)}[\rho_\Phi(x)].
\end{gather*}

Under the Helstrom measurement, Eq.~(\ref{empirical_risk_bound}) becomes

\begin{equation*}
\hat{R}(\Phi) = \frac{1}{2} -D_\mathrm{tr}(p^-\rho^-(\Phi), p^+\rho^+(\Phi))
\end{equation*}

Then $\forall \Phi \in \mathcal{F}$,

\begin{align}
\hat{R}^*(X,y;\mathcal{F})
&= \inf_{\Phi \in \mathcal{F}} \hat{R}^*(\Phi) \nonumber \\[0.5em]
&= \inf_{\Phi \in \mathcal{F}}
\left(
\frac{1}{2}
- D_{\mathrm{tr}}\!\big(
p^- \rho^-(\Phi),
\, p^+ \rho^+(\Phi)
\big)
\right) \nonumber \\[0.5em]
&= \frac{1}{2}
- \sup_{\Phi \in \mathcal{F}}
D_{\mathrm{tr}}\!\big(
p^- \rho^-(\Phi),
\, p^+ \rho^+(\Phi)
\big).
\label{loss_with_XyF}
\end{align}

For a coupling $\pi$ whose marginals are $\hat P_\pm(x)$,

\begin{align}
\rho^{+}(\Phi) - \rho^{-}(\Phi)
&= \mathbb{E}_{(x_i,x_j)\sim \pi}
\!\left[ \rho_{\Phi}(x_i) \right]
-
\mathbb{E}_{(x_i,x_j)\sim \pi}
\!\left[ \rho_{\Phi}(x_j) \right] \nonumber \\[0.5em]
&= \mathbb{E}_{(x_i,x_j)\sim \pi}
\!\left[
\rho_{\Phi}(x_i) - \rho_{\Phi}(x_j)
\right].
\label{state_diff}
\end{align}

Using Eq.~(\ref{trace_lipschitz}), Eqs.~(\ref{state_diff}), and Jensen inequality,

\begin{align}
D_{\mathrm{tr}}\!\big(\rho^{+}(\Phi), \rho^{-}(\Phi)\big)
&= \frac{1}{2}\,
\big\| \rho^{+}(\Phi) - \rho^{-}(\Phi) \big\|_{1} \nonumber \\[0.5em]
&= \frac{1}{2}\,
\Big\|
\mathbb{E}_{(x_i,x_j)\sim \pi}
\!\left[
\rho_{\Phi}(x_i) - \rho_{\Phi}(x_j)
\right]
\Big\|_{1} \nonumber \\[0.5em]
&\le
\mathbb{E}_{(x_i,x_j)\sim \pi}
\!\left[
\frac{1}{2}
\big\|
\rho_{\Phi}(x_i) - \rho_{\Phi}(x_j)
\big\|_{1}
\right] \nonumber \\[0.5em]
&=
\mathbb{E}_{(x_i,x_j)\sim \pi}
\!\left[
D_{\mathrm{tr}}\!\big(
\rho_{\Phi}(x_i),
\rho_{\Phi}(x_j)
\big)
\right] \nonumber \\[0.5em]
&\le
\mathbb{E}_{(x_i,x_j)\sim \pi}
\!\left[
\kappa_{\mathcal{F}}\,
\| x_i - x_j \|_{1}
\right] \nonumber \\[0.5em]
&=
\kappa_{\mathcal{F}}\,
\mathbb{E}_{(x_i,x_j)\sim \pi}
\!\left[
\| x_i - x_j \|_{1}
\right].
\label{long_eq}
\end{align}

The definition of 1-Wasserstein distance is
\begin{equation}
\inf_{\pi \in \Pi\!\big(\hat P_{+}(x), \hat P_{-}(x)\big)}
\mathbb{E}_{(x_i,x_j)\sim \pi}
\!\left[
\| x_i - x_j \|_{1}
\right]
=
W_{1}\!\big(
\hat P_{+}(x),
\hat P_{-}(x)
\big).
\label{was_def}
\end{equation}

Eqs.~(\ref{long_eq}) and ~(\ref{was_def}) yield Eq.~(\ref{trace_lipschitz_wasser}), as restated below.

\begin{equation*}
D_\mathrm{tr}(\rho^+(\Phi), \rho^-(\Phi)) \le \kappa_\mathcal{F} W_1(\hat{P}_+(x), \hat{P}_-(x) )
\end{equation*}

Decomposing the difference between prior-weighted class states into a state difference term and a prior difference term yields

\begin{align}
p^{+}\rho^{+}(\Phi) - p^{-}\rho^{-}(\Phi)
&=
p^{-}\!\big(\rho^{+}(\Phi) - \rho^{-}(\Phi)\big)
+
\big(p^{+} - p^{-}\big)\rho^{+}(\Phi).
\label{decompose1}
\end{align}

Using $\|\rho\|_1=\mathrm{Tr}(\rho)=1$, Eq.~(\ref{decompose1}) can be bounded in two equivalent ways,

\begin{align}
\big\| p^{+}\rho^{+}(\Phi) - p^{-}\rho^{-}(\Phi) \big\|_{1}
&\le
p^{-}\,\big\| \rho^{+}(\Phi) - \rho^{-}(\Phi) \big\|_{1} \nonumber \\
&\quad
+ \big|p^{+}-p^{-}\big|\,\big\|\rho^{+}(\Phi)\big\|_{1} \nonumber \\
\big\| p^{+}\rho^{+}(\Phi) - p^{-}\rho^{-}(\Phi) \big\|_{1}
&\le
p^{+}\,\big\| \rho^{+}(\Phi) - \rho^{-}(\Phi) \big\|_{1} \nonumber \\
&\quad
+ \big|p^{+}-p^{-}\big|\,\big\|\rho^{-}(\Phi)\big\|_{1}.
\label{decompose2}
\end{align}

Using Eqs.~(\ref{decompose2}) and (\ref{trace_lipschitz_wasser}),

\begin{align}
D_{\mathrm{tr}}\!\big(
p^{+}\rho^{+}(\Phi),
p^{-}\rho^{-}(\Phi)
\big)
&=
\frac{1}{2}
\big\|
p^{+}\rho^{+}(\Phi)
-
p^{-}\rho^{-}(\Phi)
\big\|_{1} \nonumber \\[0.3em]
&\le
\min\{p^{+},p^{-}\}
\,\frac{1}{2}
\big\|
\rho^{+}(\Phi)
-
\rho^{-}(\Phi)
\big\|_{1} \nonumber\\
&\quad
+ \frac{1}{2}\big|p^{+}-p^{-}\big| \nonumber \\[0.3em]
&=
\min\{p^{+},p^{-}\}
\,D_{\mathrm{tr}}\!\big(
\rho^{+}(\Phi),
\rho^{-}(\Phi)
\big) \nonumber\\
&\quad
+ \frac{1}{2}\big|p^{+}-p^{-}\big| \nonumber \\[0.3em]
&\le
\min\{p^{+},p^{-}\}
\,\kappa_{\mathcal{F}}\,
W_{1}\!\big(
\hat P_{+}(x),
\hat P_{-}(x)
\big) \nonumber\\
&\quad
+ \frac{1}{2}\big|p^{+}-p^{-}\big|.
\label{final_1}
\end{align}

If $p^+ = p^-$, $\min\{p^+, p^-\}=\frac{1}{2}$ and $|p^+-p^-|=0$. Then Eqs.~(\ref{final_1}) simplify to

\begin{align}
D_{\mathrm{tr}}\!\big(
p^{+}\rho^{+}(\Phi),
p^{-}\rho^{-}(\Phi)
\big)
&\le
\frac{1}{2}\,\kappa_{\mathcal{F}}\,
W_{1}\!\big(
\hat P_{+}(x),
\hat P_{-}(x)
\big).
\label{final_2}
\end{align}

Using Eqs.~(\ref{loss_with_XyF}) and ~(\ref{final_2}), the empirical loss in the embedding family $\mathcal{F}$ is bounded as

\begin{align*}
\hat{R}^{*}(X,y;\mathcal{F})
&=
\frac{1}{2}
-
\sup_{\Phi\in\mathcal{F}}
D_{\mathrm{tr}}\!\big(
p^{-}\rho^{-}(\Phi),
p^{+}\rho^{+}(\Phi)
\big) \nonumber \\[0.3em]
&\ge
\frac{1}{2}
-
\frac{1}{2}\,\kappa_{\mathcal{F}}\,
W_{1}\!\big(
\hat P_{+}(x),
\hat P_{-}(x)
\big)
\qquad \Box
\end{align*}

\section{\label{app:heatmap}Kernel-level accuracy heatmap across datasets and embeddings}

\begin{figure*}[h]
\includegraphics[scale=0.9]{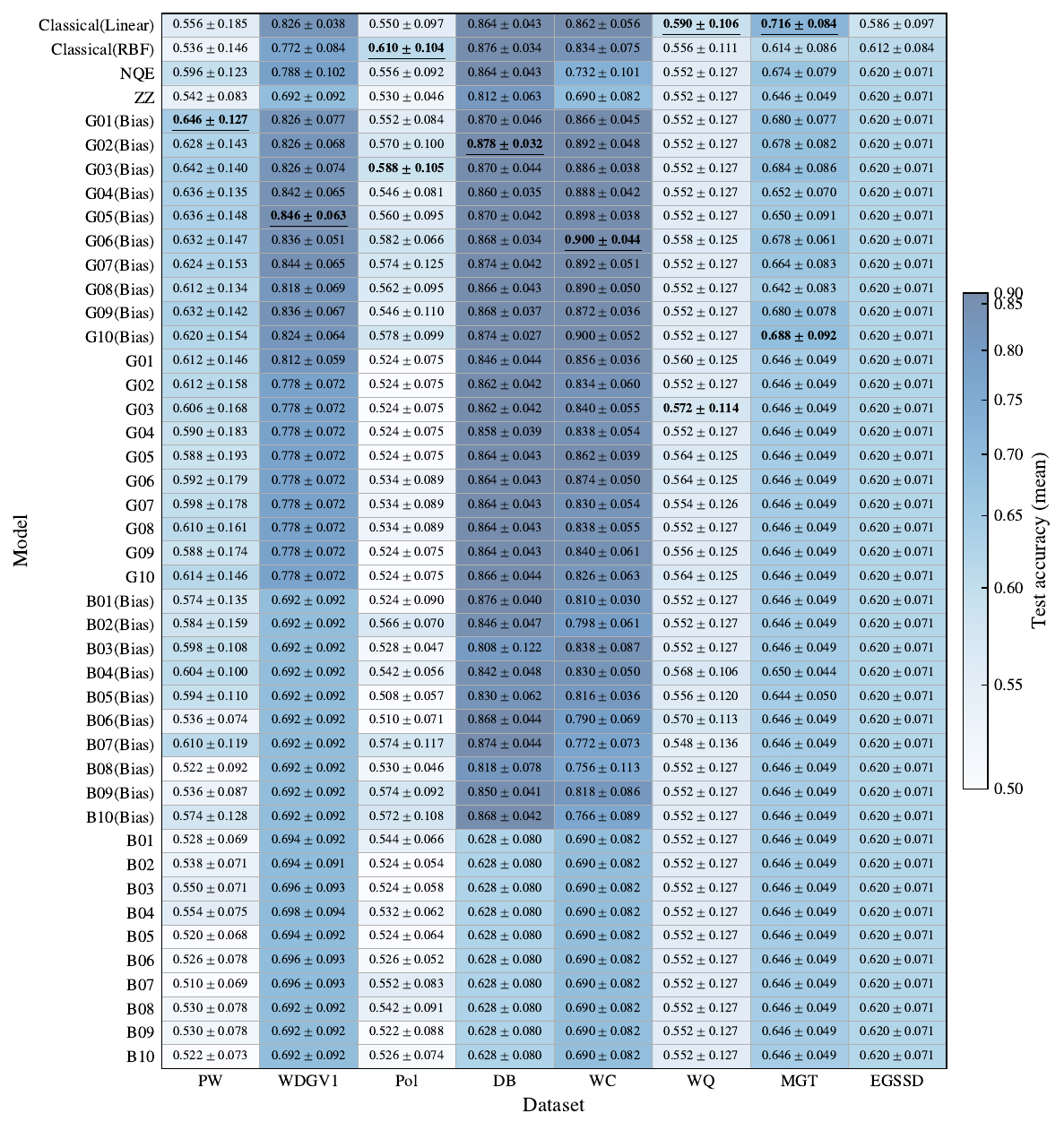}
\caption{\label{fig:indiv_heatmap} Embedding-wise QKSVM test accuracy for binary classification across datasets. Columns correspond to datasets, and rows correspond to the classical linear SVM baseline, the classical RBF SVM, the vanilla ZZ feature map, the NQE-augmented ZZ feature map, and the EGAS-generated embeddings from the $G$ and $B$ groups, with and without bias refinement. Each cell reports the mean and standard deviation of the test accuracy over ten independent train-test splits. The numerical indices $01$--$10$ label distinct EGAS-generated architectures within each group, ordered by their pre-refinement surrogate energy. Darker colors indicate higher mean test accuracy. Boldface marks the best quantum embedding within each dataset, and underlining marks the best entry among all displayed rows.
}
\end{figure*}

\clearpage

\bibliography{reference}

@article{hur2024neural,
  title={Neural quantum embedding: Pushing the limits of quantum supervised learning},
  author={Hur, Tak and Araujo, Israel F and Park, Daniel K},
  journal={Physical Review A},
  volume={110},
  number={2},
  pages={022411},
  year={2024},
  publisher={APS}
}

@InProceedings{hur2025ICML,
  title = 	 {Understanding Generalization in Quantum Machine Learning with Margins},
  author =       {Hur, Tak and Park, Daniel K.},
  booktitle = 	 {Proceedings of the 42nd International Conference on Machine Learning},
  pages = 	 {26338--26360},
  year = 	 {2025},
  editor = 	 {Singh, Aarti and Fazel, Maryam and Hsu, Daniel and Lacoste-Julien, Simon and Berkenkamp, Felix and Maharaj, Tegan and Wagstaff, Kiri and Zhu, Jerry},
  volume = 	 {267},
  series = 	 {Proceedings of Machine Learning Research},
  month = 	 {13--19 Jul},
  publisher =    {PMLR},
  url = 	 {https://proceedings.mlr.press/v267/hur25a.html}
}

@article{nakaji2024generative,
  title={The generative quantum eigensolver (GQE) and its application for ground state search},
  author={Nakaji, Kouhei and Kristensen, Lasse Bj{\o}rn and Campos-Gonzalez-Angulo, Jorge A and Vakili, Mohammad Ghazi and Huang, Haozhe and Bagherimehrab, Mohsen and Gorgulla, Christoph and Wong, FuTe and McCaskey, Alex and Kim, Jin-Sung and others},
  journal={arXiv preprint arXiv:2401.09253},
  year={2024}
}

@article{schuld2021effect,
  title={Effect of data encoding on the expressive power of variational quantum-machine-learning models},
  author={Schuld, Maria and Sweke, Ryan and Meyer, Johannes Jakob},
  journal={Physical Review A},
  volume={103},
  number={3},
  pages={032430},
  year={2021},
  publisher={APS}
}

@article{schuld2019quantum,
  title={Quantum machine learning in feature Hilbert spaces},
  author={Schuld, Maria and Killoran, Nathan},
  journal={Physical review letters},
  volume={122},
  number={4},
  pages={040504},
  year={2019},
  publisher={APS}
}

@article{havlivcek2019supervised,
  title={Supervised learning with quantum-enhanced feature spaces},
  author={Havl{\'\i}{\v{c}}ek, Vojt{\v{e}}ch and C{\'o}rcoles, Antonio D and Temme, Kristan and Harrow, Aram W and Kandala, Abhinav and Chow, Jerry M and Gambetta, Jay M},
  journal={Nature},
  volume={567},
  number={7747},
  pages={209--212},
  year={2019},
  publisher={Nature Publishing Group UK London}
}

@article{lloyd2020quantum,
  title={Quantum embeddings for machine learning},
  author={Lloyd, Seth and Schuld, Maria and Ijaz, Aroosa and Izaac, Josh and Killoran, Nathan},
  journal={arXiv preprint arXiv:2001.03622},
  year={2020}
}

@article{hubregtsen2022training,
  title={Training quantum embedding kernels on near-term quantum computers},
  author={Hubregtsen, Thomas and Wierichs, David and Gil-Fuster, Elies and Derks, Peter-Jan HS and Faehrmann, Paul K and Meyer, Johannes Jakob},
  journal={Physical Review A},
  volume={106},
  number={4},
  pages={042431},
  year={2022},
  publisher={APS}
}

@inproceedings{gentinetta2023quantum,
  title={Quantum kernel alignment with stochastic gradient descent},
  author={Gentinetta, Gian and Sutter, David and Zoufal, Christa and Fuller, Bryce and Woerner, Stefan},
  booktitle={2023 IEEE International Conference on Quantum Computing and Engineering (QCE)},
  volume={1},
  pages={256--262},
  year={2023},
  organization={IEEE}
}

@article{recio2025single,
  title={Single-shot quantum machine learning},
  author={Recio-Armengol, Erik and Eisert, Jens and Meyer, Johannes Jakob},
  journal={Physical Review A},
  volume={111},
  number={4},
  pages={042420},
  year={2025},
  publisher={APS}
}

@article{biamonte2017quantum,
  title={Quantum machine learning},
  author={Biamonte, Jacob and Wittek, Peter and Pancotti, Nicola and Rebentrost, Patrick and Wiebe, Nathan and Lloyd, Seth},
  journal={Nature},
  volume={549},
  number={7671},
  pages={195--202},
  year={2017},
  publisher={Nature Publishing Group UK London}
}

@article{cerezo2022challenges,
  title={Challenges and opportunities in quantum machine learning},
  author={Cerezo, Marco and Verdon, Guillaume and Huang, Hsin-Yuan and Cincio, Lukasz and Coles, Patrick J},
  journal={Nature computational science},
  volume={2},
  number={9},
  pages={567--576},
  year={2022},
  publisher={Nature Publishing Group US New York}
}

@article{rebentrost2014quantum,
  title={Quantum support vector machine for big data classification},
  author={Rebentrost, Patrick and Mohseni, Masoud and Lloyd, Seth},
  journal={Physical review letters},
  volume={113},
  number={13},
  pages={130503},
  year={2014},
  publisher={APS}
}

@article{schuld2017implementing,
  title={Implementing a distance-based classifier with a quantum interference circuit},
  author={Schuld, Maria and Fingerhuth, Mark and Petruccione, Francesco},
  journal={Europhysics Letters},
  volume={119},
  number={6},
  pages={60002},
  year={2017},
  publisher={IOP Publishing}
}

@article{giovannetti2008quantum,
  title={Quantum random access memory},
  author={Giovannetti, Vittorio and Lloyd, Seth and Maccone, Lorenzo},
  journal={Physical review letters},
  volume={100},
  number={16},
  pages={160501},
  year={2008},
  publisher={APS}
}

@article{kubler2021inductive,
  title={The inductive bias of quantum kernels},
  author={K{\"u}bler, Jonas and Buchholz, Simon and Sch{\"o}lkopf, Bernhard},
  journal={Advances in Neural Information Processing Systems},
  volume={34},
  pages={12661--12673},
  year={2021}
}

@article{caro2021encoding,
  title={Encoding-dependent generalization bounds for parametrized quantum circuits},
  author={Caro, Matthias C and Gil-Fuster, Elies and Meyer, Johannes Jakob and Eisert, Jens and Sweke, Ryan},
  journal={Quantum},
  volume={5},
  pages={582},
  year={2021},
  publisher={Verein zur F{\"o}rderung des Open Access Publizierens in den Quantenwissenschaften}
}

@article{thanasilp2024exponential,
  title={Exponential concentration in quantum kernel methods},
  author={Thanasilp, Supanut and Wang, Samson and Cerezo, Marco and Holmes, Zo{\"e}},
  journal={Nature communications},
  volume={15},
  number={1},
  pages={5200},
  year={2024},
  publisher={Nature Publishing Group UK London}
}

@article{giuntini2021quantum,
  title={Quantum state discrimination for supervised classification},
  author={Giuntini, Roberto and Freytes, Hector and Park, Daniel K and Blank, Carsten and Holik, Federico and Chow, Keng Loon and Sergioli, Giuseppe},
  journal={arXiv preprint arXiv:2104.00971},
  year={2021}
}

@article{helstrom1969quantum,
  title={Quantum detection and estimation theory},
  author={Helstrom, Carl W},
  journal={Journal of Statistical Physics},
  volume={1},
  number={2},
  pages={231--252},
  year={1969},
  publisher={Springer}
}

@book{schuld2021machine,
  title={Machine learning with quantum computers},
  author={Schuld, Maria and Petruccione, Francesco},
  volume={676},
  year={2021},
  publisher={Springer}
}

@article{du2022quantum,
  title={Quantum circuit architecture search for variational quantum algorithms},
  author={Du, Yuxuan and Huang, Tao and You, Shan and Hsieh, Min-Hsiu and Tao, Dacheng},
  journal={npj Quantum Information},
  volume={8},
  number={1},
  pages={62},
  year={2022},
  publisher={Nature Publishing Group UK London}
}

@article{incudini2024automatic,
  title={Automatic and effective discovery of quantum kernels},
  author={Incudini, Massimiliano and Bosco, Daniele Lizzio and Martini, Francesco and Grossi, Michele and Serra, Giuseppe and Di Pierro, Alessandra},
  journal={IEEE Transactions on Emerging Topics in Computational Intelligence},
  year={2024},
  publisher={IEEE}
}

@article{fosel2021quantum,
  title={Quantum circuit optimization with deep reinforcement learning},
  author={F{\"o}sel, Thomas and Niu, Murphy Yuezhen and Marquardt, Florian and Li, Li},
  journal={arXiv preprint arXiv:2103.07585},
  year={2021}
}

@inproceedings{altmann2024challenges,
  title={Challenges for reinforcement learning in quantum circuit design},
  author={Altmann, Philipp and Stein, Jonas and K{\"o}lle, Michael and B{\"a}rligea, Adelina and Zorn, Maximilian and Gabor, Thomas and Phan, Thomy and Feld, Sebastian and Linnhoff-Popien, Claudia},
  booktitle={2024 IEEE International Conference on Quantum Computing and Engineering (QCE)},
  volume={1},
  pages={1600--1610},
  year={2024},
  organization={IEEE}
}

@article{ostaszewski2021reinforcement,
  title={Reinforcement learning for optimization of variational quantum circuit architectures},
  author={Ostaszewski, Mateusz and Trenkwalder, Lea M and Masarczyk, Wojciech and Scerri, Eleanor and Dunjko, Vedran},
  journal={Advances in neural information processing systems},
  volume={34},
  pages={18182--18194},
  year={2021}
}

@article{pellow2024hybrid,
  title={Hybrid genetic optimization for quantum feature map design},
  author={Pellow-Jarman, Rowan and Pillay, Anban and Sinayskiy, Ilya and Petruccione, Francesco},
  journal={Quantum Machine Intelligence},
  volume={6},
  number={2},
  pages={45},
  year={2024},
  publisher={Springer}
}

@article{rapp2025reinforcement,
  title={Reinforcement learning-based architecture search for quantum machine learning},
  author={Rapp, Frederic and Kreplin, David A and Huber, Marco F and Roth, Marco},
  journal={Machine Learning: Science and Technology},
  volume={6},
  number={1},
  pages={015041},
  year={2025},
  publisher={IOP Publishing}
}

@book{nielsen2010quantum,
  title={Quantum computation and quantum information},
  author={Nielsen, Michael A and Chuang, Isaac L},
  year={2010},
  publisher={Cambridge university press}
}

@article{huang2021power,
  title={Power of data in quantum machine learning},
  author={Huang, Hsin-Yuan and Broughton, Michael and Mohseni, Masoud and Babbush, Ryan and Boixo, Sergio and Neven, Hartmut and McClean, Jarrod R},
  journal={Nature communications},
  volume={12},
  number={1},
  pages={2631},
  year={2021},
  publisher={Nature Publishing Group UK London}
}

@article{bae2015quantum,
  title={Quantum state discrimination and its applications},
  author={Bae, Joonwoo and Kwek, Leong-Chuan},
  journal={Journal of Physics A: Mathematical and Theoretical},
  volume={48},
  number={8},
  pages={083001},
  year={2015},
  publisher={IOP Publishing}
}

@article{Lee2024QST,
doi = {10.1088/2058-9565/ad0a05},
url = {https://doi.org/10.1088/2058-9565/ad0a05},
year = {2023},
month = {nov},
publisher = {IOP Publishing},
volume = {9},
number = {1},
pages = {015017},
author = {Lee, Dongkeun and Baek, Kyunghyun and Huh, Joonsuk and Park, Daniel K},
title = {Variational quantum state discriminator for supervised machine learning},
journal = {Quantum Science and Technology},
}

@article{buhrman2001quantum,
  title={Quantum fingerprinting},
  author={Buhrman, Harry and Cleve, Richard and Watrous, John and De Wolf, Ronald},
  journal={Physical review letters},
  volume={87},
  number={16},
  pages={167902},
  year={2001},
  publisher={APS}
}

@book{wilde2013quantum,
  title={Quantum information theory},
  author={Wilde, Mark M},
  year={2013},
  publisher={Cambridge university press}
}

@article{fuchs2002cryptographic,
  title={Cryptographic distinguishability measures for quantum-mechanical states},
  author={Fuchs, Christopher A and Van De Graaf, Jeroen},
  journal={IEEE Transactions on Information Theory},
  volume={45},
  number={4},
  pages={1216--1227},
  year={2002},
  publisher={IEEE}
}

@article{wang2023fast,
  title={Fast quantum algorithms for trace distance estimation},
  author={Wang, Qisheng and Zhang, Zhicheng},
  journal={IEEE Transactions on Information Theory},
  volume={70},
  number={4},
  pages={2720--2733},
  year={2023},
  publisher={IEEE}
}

@article{kolle2024reinforcement,
  title={A reinforcement learning environment for directed quantum circuit synthesis},
  author={K{\"o}lle, Michael and Schubert, Tom and Altmann, Philipp and Zorn, Maximilian and Stein, Jonas and Linnhoff-Popien, Claudia},
  journal={arXiv preprint arXiv:2401.07054},
  year={2024}
}

@article{kundu2024enhancing,
  title={Enhancing variational quantum state diagonalization using reinforcement learning techniques},
  author={Kundu, Akash and Bede{\l}ek, Przemys{\l}aw and Ostaszewski, Mateusz and Danaci, Onur and Patel, Yash J and Dunjko, Vedran and Miszczak, Jaros{\l}aw A},
  journal={New Journal of Physics},
  volume={26},
  number={1},
  pages={013034},
  year={2024},
  publisher={IOP Publishing}
}

@phdthesis{montagna2021quantum,
  title={Quantum circuit design with reinforcement learning},
  author={Montagna, Francesco},
  year={2021},
  school={Politecnico di Torino}
}

@article{islam2025quantum,
  title={Quantum Circuit Synthesis Using Fuzzy-Logic-Assisted Genetic Algorithms},
  author={Islam, Ishraq and Jha, Vinayak and Thomas, Sneha and Egan, Kieran F and Nobel, Alvir and Kim, Serom and Chaudhary, Manu and Ogundele, Sunday and Kneidel, Dylan and Phillips, Ben and others},
  journal={Algorithms},
  volume={18},
  number={4},
  pages={178},
  year={2025},
  publisher={MDPI}
}

@inproceedings{phalak2025optimizing,
  title={Optimizing Quantum Embedding using Genetic Algorithm for QML Applications},
  author={Phalak, Koustubh and Ghosh, Archisman and Ghosh, Swaroop},
  booktitle={2025 26th International Symposium on Quality Electronic Design (ISQED)},
  pages={1--9},
  year={2025},
  organization={IEEE}
}

@article{radford2019language,
  title={Language models are unsupervised multitask learners},
  author={Radford, Alec and Wu, Jeffrey and Child, Rewon and Luan, David and Amodei, Dario and Sutskever, Ilya and others},
  journal={OpenAI blog},
  volume={1},
  number={8},
  pages={9},
  year={2019}
}

@misc{waveform_database_generator_version_1_107,
  author       = {Breiman, L. and Stone, C.J.},
  title        = {{Waveform Database Generator (Version 1)}},
  year         = {1984},
  howpublished = {UCI Machine Learning Repository},
  note         = {{DOI}: https://doi.org/10.24432/C5CS3C}
}

@article{vanschoren2014openml,
  title={OpenML: networked science in machine learning},
  author={Vanschoren, Joaquin and Van Rijn, Jan N and Bischl, Bernd and Torgo, Luis},
  journal={ACM SIGKDD Explorations Newsletter},
  volume={15},
  number={2},
  pages={49--60},
  year={2014},
  publisher={ACM New York, NY, USA}
}

@article{grinsztajn2022tree,
  title={Why do tree-based models still outperform deep learning on typical tabular data?},
  author={Grinsztajn, L{\'e}o and Oyallon, Edouard and Varoquaux, Ga{\"e}l},
  journal={Advances in neural information processing systems},
  volume={35},
  pages={507--520},
  year={2022}
}

@misc{phishing_websites_327,
  author       = {Mohammad, Rami and McCluskey, Lee},
  title        = {{Phishing Websites}},
  year         = {2012},
  howpublished = {UCI Machine Learning Repository},
  note         = {{DOI}: https://doi.org/10.24432/C51W2X}
}

@misc{magic_gamma_telescope_159,
  author       = {Bock, R.},
  title        = {{MAGIC Gamma Telescope}},
  year         = {2004},
  howpublished = {UCI Machine Learning Repository},
  note         = {{DOI}: https://doi.org/10.24432/C52C8B}
}

@misc{electrical_grid_stability_simulated_data__471,
  author       = {Arzamasov, Vadim},
  title        = {{Electrical Grid Stability Simulated Data }},
  year         = {2018},
  howpublished = {UCI Machine Learning Repository},
  note         = {{DOI}: https://doi.org/10.24432/C5PG66}
}

@misc{dry_bean_602,
  title        = {{Dry Bean}},
  year         = {2020},
  howpublished = {UCI Machine Learning Repository},
  note         = {{DOI}: https://doi.org/10.24432/C50S4B}
}

@misc{wine_quality_186,
  author       = {Cortez, Paulo and Cerdeira, A. and Almeida, F. and Matos, T. and Reis, J.},
  title        = {{Wine Quality}},
  year         = {2009},
  howpublished = {UCI Machine Learning Repository},
  note         = {{DOI}: https://doi.org/10.24432/C56S3T}
}

@article{bergholm2018pennylane,
  title={Pennylane: Automatic differentiation of hybrid quantum-classical computations},
  author={Bergholm, Ville and Izaac, Josh and Schuld, Maria and Gogolin, Christian and Ahmed, Shahnawaz and Ajith, Vishnu and Alam, M Sohaib and Alonso-Linaje, Guillermo and AkashNarayanan, B and Asadi, Ali and others},
  journal={arXiv preprint arXiv:1811.04968},
  year={2018}
}

@article{jensen1906fonctions,
  title={Sur les fonctions convexes et les in{\'e}galit{\'e}s entre les valeurs moyennes},
  author={Jensen, Johan Ludwig William Valdemar},
  journal={Acta mathematica},
  volume={30},
  number={1},
  pages={175--193},
  year={1906},
  publisher={Springer}
}

@book{villani2008optimal,
  title={Optimal transport: old and new},
  author={Villani, C{\'e}dric and others},
  volume={338},
  year={2008},
  publisher={Springer}
}

@article{furrutter2025synthesis,
  title={Synthesis of discrete-continuous quantum circuits with multimodal diffusion models},
  author={F{\"u}rrutter, Florian and Chandani, Zohim and Hamamura, Ikko and Briegel, Hans J and Mu{\~n}oz-Gil, Gorka},
  journal={arXiv preprint arXiv:2506.01666},
  year={2025}
}

@incollection{efron1992bootstrap,
  title={Bootstrap methods: another look at the jackknife},
  author={Efron, Bradley},
  booktitle={Breakthroughs in statistics: Methodology and distribution},
  pages={569--593},
  year={1992},
  publisher={Springer}
}

@inproceedings{bengio2009curriculum,
  title={Curriculum learning},
  author={Bengio, Yoshua and Louradour, J{\'e}r{\^o}me and Collobert, Ronan and Weston, Jason},
  booktitle={Proceedings of the 26th annual international conference on machine learning},
  pages={41--48},
  year={2009}
}

@article{seong2025hamiltonian,
  title={Hamiltonian formulations of centroid-based clustering},
  author={Seong, Myeonghwan and Park, Daniel Kyungdeock},
  journal={Frontiers in Physics},
  volume={13},
  pages={1544623},
  year={2025},
  publisher={Frontiers Media SA}
}

@article{choi2025early,
  title={Early-stage detection of cognitive impairment by hybrid quantum-classical algorithm using resting-state functional MRI time-series},
  author={Choi, Junggu and Hur, Tak and Park, Daniel K and Shin, Na-Young and Lee, Seung-Koo and Lee, Hakbae and Han, Sanghoon},
  journal={Knowledge-Based Systems},
  volume={310},
  pages={112922},
  year={2025},
  publisher={Elsevier}
}

@article{liu2025neural,
  title={Neural quantum embedding via deterministic quantum computation with one qubit},
  author={Liu, Hongfeng and Hur, Tak and Zhang, Shitao and Che, Liangyu and Long, Xinyue and Wang, Xiangyu and Huang, Keyi and Fan, Yu-ang and Zheng, Yuxuan and Feng, Yufang and others},
  journal={Physical Review Letters},
  volume={135},
  number={8},
  pages={080603},
  year={2025},
  publisher={APS}
}

@article{kim2025multi,
  title={Multi-Channel Convolutional Neural Quantum Embedding},
  author={Kim, Yujin and Im, Changjae and Kim, Taehyun and Hur, Tak and Park, Daniel K},
  journal={Advanced Quantum Technologies},
  pages={e00575},
  year={2025},
  publisher={Wiley Online Library}
}

@Article{Kim2023Nat,
author={Kim, Youngseok
and Eddins, Andrew
and Anand, Sajant
and Wei, Ken Xuan
and van den Berg, Ewout
and Rosenblatt, Sami
and Nayfeh, Hasan
and Wu, Yantao
and Zaletel, Michael
and Temme, Kristan
and Kandala, Abhinav},
title={Evidence for the utility of quantum computing before fault tolerance},
journal={Nature},
year={2023},
month={Jun},
day={01},
volume={618},
number={7965},
pages={500-505},
issn={1476-4687},
doi={10.1038/s41586-023-06096-3},
url={https://doi.org/10.1038/s41586-023-06096-3}
}

\end{document}